\documentclass[aps,ajp,nofootinbib,preprint,12pt,altaffilletter,showpacs]{revtex4}
\usepackage{graphicx}
\usepackage{amssymb}
\usepackage{amsmath} 
\usepackage{hyperref}
\usepackage{textcomp}
\usepackage{psfrag}
\usepackage{setspace}
\usepackage{url}
\usepackage{bm}
\usepackage{subfigure}
\begin{document}
\title{Rheology in the Teaching Lab: Properties of Starch Suspensions}
\author{Joel~A.~Groman}
\author{James~G.~Miller}
\affiliation{Department of Physics, Washington University, St.~Louis,
Mo.~63130}
\author{Jonathan~I.~Katz$^*$}
\affiliation{Department of Physics}
\affiliation{McDonnell Center for the Space Sciences, Washington
University, St.~Louis, Mo.~63130}
\email{katz@wuphys.wustl.edu}
\date{\today}
\begin{abstract}
In everyday life we encounter many complex fluids, from shear-thinning paint
and toothpaste to shear-thickening starch suspensions.  The study of their
properties offers an opportunity for students to relate sophisticated
physical concepts to their everyday experience.  Modern rheology uses
expensive equipment impractical for the teaching laboratory.  Here we
describe a rudimentary rheometer suitable for student laboratories that can
demonstrate and quantify discontinuous shear thickening, the most dramatic
property of complex fluids, and use it to measure the properties of starch
suspensions.
\end{abstract}
\maketitle
\section{Introduction}
Simple fluids, such as water, honey, oils, pitch and liquid nitrogen have
the ``Newtonian'' property that their stress is proportional to their strain
rate (flow rate).  Their ratio is a scalar viscosity\footnote{This
discussion is limited to essentially incompressible fluids, usually an
excellent approximation.  When compressibility is important, as for sound
waves, an additional scalar bulk viscosity must be defined.}.  This
proportionality defines a simple fluid, whatever the value of the viscosity.

In contrast, there are ``non-Newtonian'' fluids with more complicated
relations between stress and strain rate.  These fluids may contain
polymers, or be suspensions (typically in a Newtonian solvent) of solid
particles, membrane-bound vesicles or droplets of an immiscible fluid
(emulsions).

Paint, ketchup, toothpaste and corn starch suspensions are familiar examples
of non-Newtonian fluids.  Most of these are shear-thinning: they may have a
small finite strength at rest (which is why toothpaste doesn't flow out of
its tube unless squeezed, or ketchup out of its bottle, unless squeezed,
shaken or struck) or a viscosity that decreases as the flow rate is
increased (so that paint is easily spread with a brush, but doesn't drip
once spread).  Unlike these, starch suspensions have the remarkable
property, known to schoolchildren who gave them the nickname ``oobleck''
after a fictional substance, of suddenly turning stiff, increasing their
viscosity by orders of magnitude, if the strain rate exceeds a threshold.
Brown and Jaeger \cite{BJ13} provide a recent review of the properties of
these suspensions.

A student laboratory experiment will excite more interest if it is novel,
if it explores a dramatic phenomenon, and if it is related to students'
everyday experience.  The ``discontinuous'' (abrupt) shear stiffening of
starch suspensions meets these criteria.  Yet quantitative rheometry
requires expensive and delicate equipment unavailable in and unsuitable for
the student laboratory.  Here we describe, and report on results obtained
with, a rudimentary rheometer that can be assembled from a few dollars'
worth of equipment.  With the aid of a consumer-grade video camera, it can
produce quantitative data.
\section{The Instrument}
\begin{figure}
\begin{center}
\includegraphics[width=3in]{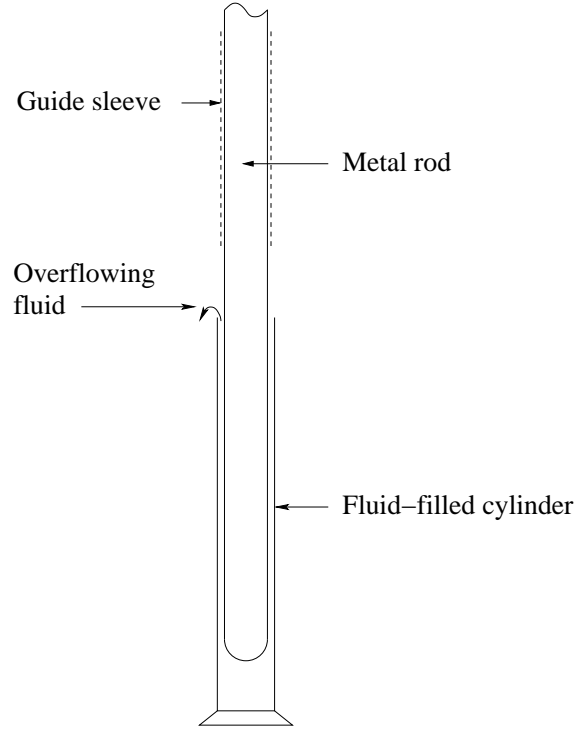}
\end{center}
\caption{\label{gromanexpt}The rudimentary rheometer consists of a cylinder
filled with the fluid whose properties are to be measured.  A metal rod with
diameter slightly less than that of the cylinder sinks into the fluid,
driving fluid up the annulus between rod and cylinder.  The rod is guided
and centered in the cylinder by a cylindrical sleeve aligned with the
cylinder axis.}
\end{figure}

The instrument is shown, approximately to scale, in Fig.~\ref{gromanexpt}.
A light-emitting diode mounted on the top of the rod was used to determine
the position of the rod against an aligned meter stick with an attached LED
that serves as a reference.  Data were recorded with a video camera at one
or 30 frames per second (the lower recording rate was used for more slowly
sinking rods because of limited camera memory), and the velocity averaged
over 20 frames if recorded at 30 fps and over 5 frames if recorded at 1 fps.
Averaging was necessary because the rod position was determined to only $\pm
0.5$ line in the video image, or about $\pm 0.1$ mm; accuracy was limited by
the resolution of the video image.  Data were processed with ImageJ software
\cite{imagej}.  The rods were 36 cm long, rounded to a hemisphere at their
lower ends, with diameters 18.9 mm.  The guide sleeve had an internal
diameter of 19.8 mm and was 12.5 cm long.  The cylinder (a nominal 50 ml
graduated cylinder but with larger total volume) had an internal diameter
of 23.5 mm and depth, rim to interior bottom, of 16.5 cm.  The aluminum rod
had a mass of 271 g and the stainless steel rod a mass of 820 g.

\section{Theory}
\subsection{Newtonian Fluids}
We make the approximation that the annulus, of uniform width $h$ (if the
rod is perfectly centered in the cylinder) is thin compared to the radius
$r$ of the rod.  Then the geometry of the annulus may be approximated as
that of a planar duct.  The full theory \cite{M36} of the flow in a
cylindrical annular duct in which the inner wall is moving with respect to
the outer wall is cumbersome, and its use would not be justified in these
experiments in which the geometry cannot be controlled precisely and the
two cylinders may not be accurately coaxial.  If $h \ll r$ the rod sinks 
much more slowly that the fluid flows, so that the solution for flow in a
planar duct with stationary walls may be used.  Because the gap is
everywhere narrow compared to $r$ the case of an off-center rod is also
readily dealt with.

For a Newtonian fluid with single-valued dynamic viscosity $\eta$ and a
pressure gradient $dp \over dz$ parallel to the duct walls the fluid
velocity
\begin{equation}
\label{ductn}
v(y) = {1 \over 2 \eta}{dp \over dz}\left(y^2 - {h^2 \over 4}\right),
\end{equation}
where $y$ is the transverse coordinate in the duct with $y = 0$ at its
midplane.  If the rod is centered in the cylinder the fluid flow rate per
unit circumference
\begin{equation}
\label{q}
{\dot q} = \int^{h/2}_{-h/2} v(y)\,dy = -{h^3 \over 12 \eta}
{dp \over dz}
\end{equation}
and the total fluid flow rate ${\dot Q} = 2 \pi r {\dot q}$.  The rod sinks
at a rate 
\begin{equation}
\label{vrod}
v_{rod} = {{\dot Q} \over \pi r^2} = {h^3 \over 6 \eta r}{dp \over dz}.
\end{equation}
The ratio of $v_{rod}$ to the fluid velocity at $y = 0$
\begin{equation}
\label{vrod2}
\left\vert{v_{rod} \over v(0)}\right\vert = {4 \over 3}{h \over r} \ll 1,
\end{equation}
justifying the use of the solution Eq.~\ref{ductn} for flow in a duct with
stationary walls.

The pressure gradient in the flowing fluid is given by the weight per unit
cross-sectional area of the rod, allowing for the buoyancy of the displaced
fluid, divided by its immersed length $z_r$:
\begin{equation}
{dp \over dz} = {g(\rho L - \rho_f z_r) \over z_r},
\end{equation}
where $\rho$ is the density of the rod (2.7 g/cm$^3$ for aluminum, 8.0
gm/cm$^3$ for stainless steel) and $\rho_f$ is the density of the fluid.
The buoyancy correction $\rho_f z_r$ is small if the immersed length is
small compared to the rod length, as it will always be when the rod has
just entered the fluid, and is never more than about 20\% in our
experiments.

In order to obtain a simple analytic result we neglect buoyancy.  The
equation of motion of the rod (inertia is negligible for a viscous fluid) is
\begin{equation}
\label{zr}
{d z_r \over dt} = v_{rod} = {h^3 g \rho L \over 6 \eta r z_r}
\end{equation}
with the elementary solution
\begin{equation}
\label{vn}
v_{rod} = \sqrt{h^3 g \rho L \over 12 \eta r} t^{-1/2}.
\end{equation}

Although we use a guide tube to try to keep the sinking rod as close and
parallel to the axis of the fluid-filled cylinder as possible, alignment
is not perfect and we consider the effects of its being off-center.  If
the axis of the rod is displaced from the axis of the cylinder by $\Delta x$
the width of the gap between rod and cylinder, to lowest order in the small
quantity $\Delta x/r$ ($\Delta x \le h$), is
\begin{equation}
\Delta r(\theta) \approx h - \Delta x \cos{\theta},
\end{equation}
where $\theta$ is the angle from the direction of $\vec {\Delta x}$.  Then
\begin{equation}
\label{offcentern}
{\dot Q} = \int_0^{2\pi} \! d\theta \, r {\dot q}(\theta) = - {\pi r h^3
\over 6 \eta} {dp \over dz} \left[1 + {3 \over 2}\left({\Delta x \over h}
\right)^2 \right].
\end{equation}

For an off-center rod $\dot Q$ and $v_{rod} = {\dot Q}/\pi r^2$ can be as
much as 5/2 times greater than for a centered rod (in the $h \ll r$
approximation).  We have ignored the fact that if the rod is very close
($\Delta r(\theta) \lesssim h \sqrt{h/r}$) to  the cylinder wall the
approximation Eq.~\ref{vrod2} is not valid and additional drag is
contributed by the relative motion of rod and wall.
\subsection{Shear stiffening fluids}
In a shear stiffening fluid the viscosity is an increasing function of the
strain rate.  The behavior of such fluids is complex, but often may be
approximated by the condition that the viscosity increases abruptly by
orders of magnitude if the strain rate $\vert {\dot \gamma} \vert > {\dot
\gamma}_c$, where ${\dot \gamma}_c$ is a critical strain rate \cite{BJ13}.
As a result, Eq.~\ref{ductn} breaks down if it implies $\vert {\dot \gamma}
\vert = {1 \over \eta} \vert {dp \over dz} y \vert > {\dot \gamma}_c$, where
$\eta$ is the viscosity in the unstiffened regime.  The value of ${\dot
\gamma}_c$ is generally taken as an empirical parameter, but has been
explained as the result of surface tension \cite{C14}.

If Eq.~\ref{ductn} implies $\vert {\dot \gamma} \vert > {\dot \gamma}_c$ the
suspension undergoes discontinuous shear thickening in the outside of the
duct, where $\vert y \vert > {\dot \gamma}_c \eta /\vert {dp \over dz}
\vert$.  This stiffening, increasing $\eta$ by orders of magnitude under
conditions in which $\eta {d^2 v \over dy^2} = {dp \over dz}$ is continuous
across the duct, implies ${dv \over dy} \to \mathrm{Constant}$, except for a
central region in which $\vert y \vert < {\dot \gamma}_c \eta / \vert {dp
\over dz}\vert$ and Eq.~\ref{ductn} applies.  However, in the limit $h/r \to
0$ the difference in velocity of the duct walls $\vert v_{rod} \vert \to 0$
(because $\vert v_{rod}/v(y) \vert = {\cal O}(h/r)$; Eq.~\ref{vrod2}) so
that the condition that the fluid velocity at a duct wall equal the wall
velocity implies $\mathrm{Constant} \to 0$ and ${\dot \gamma} \to 0$.  The
continuity of stress across the duct implies an additional factor of the
large shear thickened $\eta$ in the denominator of ${\dot \gamma} = dv/dy$
(Eq.~\ref{ductn}).  This shear-thickened solution is not self-consistent:
shear thickening reduces the strain rate below the shear-thickening 
threshold.

A self-consistent solution is found if $\vert {\dot \gamma} \vert$ remains
at the shear thickening threshold ${\dot \gamma}_c$ across the duct, aside
from the central region \cite{LBS91}.  Then, assuming this central region
is negligibly thin, 
\begin{equation}
{\dot q} = {\dot \gamma}_c {h^2 \over 4},
\end{equation}
\begin{equation}
{\dot Q} = {\dot \gamma}_c {\pi r h^2 \over 2}
\end{equation}
and
\begin{equation}
\label{vrodt}
v_{rod} = {{\dot Q} \over \pi r^2} = {\dot \gamma}_c {h^2 \over 2 r}.
\end{equation}
The sink rate is predicted to be independent of the weight of the rod.  It
depends on the nature of the suspension through the empirical ${\dot
\gamma}_c$ (Chu, {\it et al.\/} \cite{C14} have suggested ${\dot \gamma}_c$
depends on surface interactions and the width of the duct, but this has not
been demonstrated).

For an off-center rod the result Eq.~\ref{offcentern} is replaced by
\begin{equation}
\label{offcentert}
{\dot Q} = \int_0^{2\pi}\,d\theta r {\dot q}(\theta) = {\dot \gamma}_c
{\pi r h^2 \over 2} \left[1 + {1 \over 2}\left({\Delta x \over h}\right)^2
\right].
\end{equation}
Then $\dot Q$ and $v_{rod} = {\dot Q}/\pi r^2$ can be as
much as 3/2 times greater than for a centered rod (in the $h \ll r$
approximation).
\section{Results}
\begin{figure}
\begin{center}
\includegraphics[width=3in]{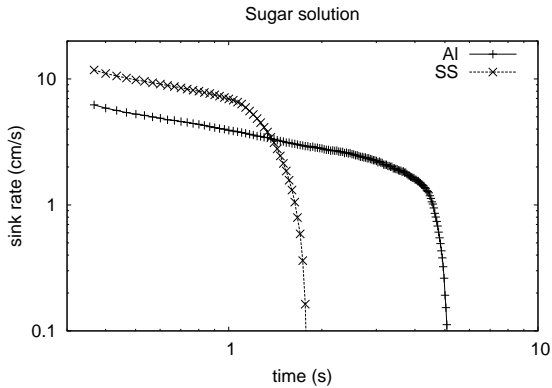}
\end{center}
\caption{\label{sugar}Sinking rate $v_{rod}$ of aluminum and stainless steel
rods in viscous cane sugar solution.  From the measured sink rate and
Eq.~\ref{vn} the viscosity $\eta \approx 0.7$ Pa-s and the Reynolds number
$\mathrm{Re} \approx 0.1$ for the Al rod and $\mathrm{Re} \approx 0.3$ for
the steel rod at $t = 1\,$s.  The data sampling rate was 30/s, but the
points shown represent boxcar averages of 20 points, taken to smooth
otherwise noisy data.}
\end{figure}

As a test of the method and apparatus, we first used a viscous solution
of cane sugar in water.  The results are shown in Fig.~\ref{sugar}.  The
proportionality $v_{rod} \propto t^{-1/2}$ for time $t$ after the rod enters
the solution agrees with the prediction Eq.~\ref{vn} for a Newtonian fluid.
The neglect of inertia in Eq.~\ref{zr} is justified by the self-consistent
result that the Reynolds number $\mathrm{Re} = h v_{rod}/\eta \ll 1$
throughout the run ($\mathrm{Re} \approx 0.1$ for the Al rod at $t = 1\,$s).

\begin{figure}
\begin{center}
\includegraphics[width=3in]{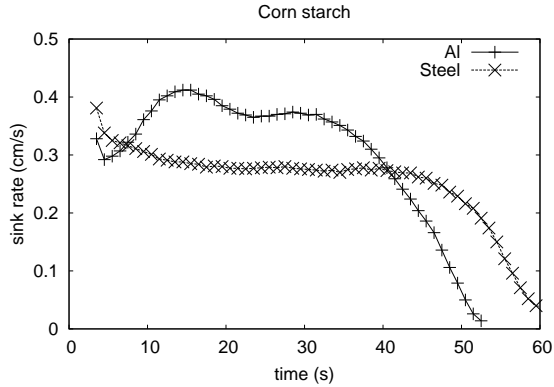}
\end{center}
\caption{\label{corn}Sink rates of aluminum and steel rods in a 43\%
suspension of corn starch.  Data were sampled every second, but each point
shown is a boxcar average over five seconds.} 
\end{figure}

\begin{figure}
\begin{center}
\includegraphics[width=3in]{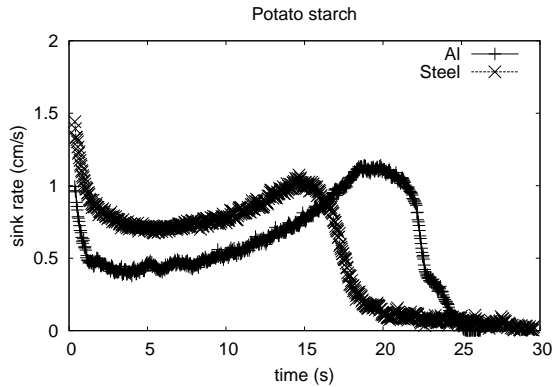}
\end{center}
\caption{\label{potato}Sink rates of aluminum and steel rods in a 43\%
suspension of potato starch.  The data sampling rate was 30/s, but the points
shown are boxcar averages over 20 samples.}
\end{figure}

\begin{figure}
\begin{center}
\includegraphics[width=3in]{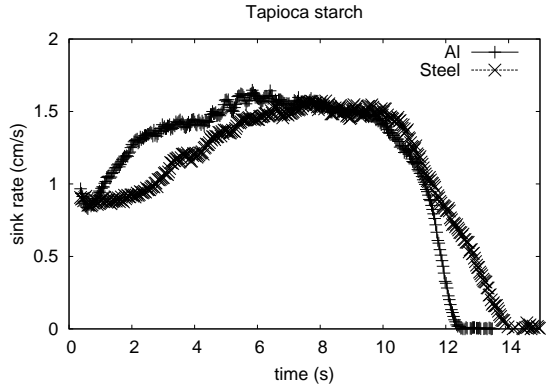}
\end{center}
\caption{\label{tapioca}Sink rates of aluminum and steel rods in a 43\%
suspension of tapioca starch.  The data were sampled at a rate of 30/s, but
each point shown is a boxcar average over 20 samples.}
\end{figure}

The sinking rates of aluminum and stainless steel rods in suspensions of
corn, potato and tapioca starches in isopycnic (density matched) CsCl brines
are shown in Figs.~\ref{corn}--\ref{tapioca}.  All suspensions had starch
volume and mass fractions of 43\%, well into the regime in which
discontinuous shear thickening occurs, but a low enough concentration that
the suspensions are shear thinning fluids (rather than pastes with finite
strength) at low strain rates.

The results for the starch suspensions are mixed.  In corn starch
(Fig.~\ref{corn}) the steel rod sank at a nearly constant rate, as expected
(Eq.~\ref{vrodt}) for a shear thickened suspension.  The implied stiffening
threshold ${\dot \gamma}_c \approx 4$/s, typical of previous measurements of
corn starch suspensions \cite{BJ13,C14,F08} (that are widely scattered,
perhaps as a result of differing properties of this poorly standardized
natural product).  The varying sink rate of the aluminum rod might be
attributed to a varying displacement from a centered position in the
cylinder (Eq.~\ref{offcentert}).  The greater stability of the sink rate of
the steel rod was observed in two other pairs of runs (not shown).

The generally increasing sink rates of potato starch suspensions may be 
attributed to motion of the rods from centered to off-center positions in
the cylinder, and the initial decrease may be the result of a transient 
phase in which the suspension is unstiffened, as expected and found for a
Newtonian fluid shown in Fig.~\ref{sugar}.  The steel rod sank about 50\%
faster than the aluminum rod through most of its descent, although towards
the end the aluminum rod speeded up to a sink rate as fast as the maximum
sink rate of the steel rod.  The inferred ${\dot \gamma}_c \approx$ 20--30/s.

The tapioca starch suspensions are close to the predictions of
Eq.~\ref{vrodt}: The sink rates were roughly constant after an initial
increase by a factor of 1.5, consistent with motion of the rods from
centered to near-wall positions (Eq.~\ref{offcentert}), and were nearly
independent of the weight of the rod.  The implied ${\dot \gamma}_c \approx
40$/s.

Qualitatively, the results for starch suspensions followed predictions:
the sink rates were roughly independent of the rod weight, in contrast to a
Newtonian fluid in which the sink rate would be proportional to the rod
mass, allowing for buoyancy, which is at least three times greater for
steel.

The results are summarized in the Table.  There is no apparent correlation
between the critical strain rates for discontinuous shear stiffening and the
size of the starch grains.
\begin{table}
\begin{center}
\begin{tabular}{|c|c|c|c|}
\hline
starch & grain diameter & ${\dot \gamma}_c$ & CsCl fraction \\
\hline
corn & 14$\mu$ & 4/s & 52.5\% \\
tapioca & 14$\mu$ & 40/s & 52.5\% \\
potato & 35$\mu$ & 20--30/s & 54.5\% \\
\hline
\end{tabular}
\end{center}
\caption{Critical strain rates ${\dot \gamma}_c$ for suspensions of three
starches in CsCl brine.  Mean grain diameters from Ref.~\cite{S84}.  The
mass fractions of CsCl in an isopycnic brine, used to prevent sedimentation,
are also shown; we found slightly different densities for the different
starches, but these values may be different for different samples of these
natural products.}
\end{table}
\section{Discussion}
We have demonstrated a simple rheometer that can be built, or used, by
students in an advanced laboratory course at negligible expense and without
requiring special facilities.  This rheometer can demonstrate basic but
unfamiliar properties of Newtonian fluid flow as well as obtaining
significant novel data about the properties of complex fluids.  It is
suitable both as a teaching tool in a curriculum that includes hydrodynamics
or rheology and as an introduction to research that produces non-trivial
results without the use of expensive state-of-the-art apparatus.
\begin{acknowledgments}
This work was supported in part by American Chemical Society Petroleum
Research Fund Grant \#51987-ND9.
\end{acknowledgments}


\begin{thebibliography}{9}
\bibitem{BJ13}E.~Brown and H.~M.~Jaeger, {\it Rep. Prog. Phys.\/} {\bf 77},
046602 (2014).
\bibitem{imagej}\url{http://imagej.nih.gov/ij/} accessed June 3, 2014.
\bibitem{M36}W.~M\"uller, {\it Zeit. angew. Math. Mech.\/} {\bf 16}, 227 
(1936).
\bibitem{C14}C.~Chu, {\it et al.\/}, Phys. Rev. E submitted
(arXiv:1405.7233) (2014).
\bibitem{LBS91}H.~M.~Laun, R.~Bung and F.~Schmidt, {\it J. Rheol.\/} {\bf
35}(6), 999 (1991).
\bibitem{F08}A.~Fall, H.~Huang, F.~Bertrand, G.~Ovarlez and D.~Bonn,
{\it Phys. Rev. Lett.\/} {\bf 100}, 018301 (2008). 
\bibitem{S84}E.~M.~Snyder, Industrial Microscopy of Starches, in {\it
Starch: Chemistry and Technology\/}, eds. R.~L.~Whistler, J.~N.~BeMiller
and E.~F.~Paschall (Academic Press, Orlando, 1984).
\end{thebibliography}
\end{document}